\documentclass[a4paper,11pt]{article}
\usepackage{pos}
\graphicspath{ {figures/} }
\newcommand{\MSb}{\overline{\text{MS}}}
\newcommand{\numax}{\nu_{\textrm{max}}}
\newcommand{\zmax}{z_{\textrm{max}}}

\title{
Unveiling generalized parton distributions through the pseudo-distribution approach}
\ShortTitle{Unveiling GPDs through the pseudo-distribution approach}

\author*[a]{Niilo Nurminen}
\author[b]{Shohini Bhattacharya}
\author[a]{Wojciech Chomicki}
\author[a]{Krzysztof Cichy}
\author[c]{Martha Constantinou}
\author[c]{Andreas Metz}
\author[d]{Fernanda Steffens}

\affiliation[a]{Faculty of Physics, Adam Mickiewicz University,\\
ul. Uniwersytetu Poznańskiego 2, 61-614 Poznań, Poland}

\affiliation[b]{RIKEN BNL Research Center,
Upton, NY 11973-5000, USA}

\affiliation[c]{Department of Physics, Temple University,\\
1925 N. 12th Street, Philadelphia, PA 19122-1801, USA}

\affiliation[d]{Institut f\"ur Strahlen- und Kernphysik, Rheinische Friedrich-Wilhelms-Universit\"at Bonn,\\
Nussallee 14-16, 53115 Bonn, Germany}

\emailAdd{niilo.nurminen@amu.edu.pl}

\abstract{
Understanding the intricate three-dimensional internal structure of the nucleon has been a long-standing challenge. The main quantitative tool to map out this structure are the generalized parton distributions (GPDs). In these proceedings, we present our exploratory results from applying the pseudo-distribution approach on the lattice, showing the unpolarized isovector $E$ GPD as an example. We use one ensemble of $N_f=2+1+1$ twisted mass fermions with a clover term, at a non-physical pion mass of $260 \textrm{ MeV}$ and a lattice spacing of $0.093 \textrm{ fm}$.}

\FullConference{The 40th International Symposium on Lattice Field Theory (Lattice 2023)\\
July 31st - August 4th, 2023\\
Fermi National Accelerator Laboratory\\}

\begin{document}
\maketitle

\section{Introduction}
The precise structure of the most fundamental baryon, the nucleon, which is the building block of all matter, still eludes our understanding. Unraveling the composition of hadrons has been a persistent quest for all particle physicists, theoreticians and experimentalists alike, for decades. The partonic description of nucleons in the form of parton distribution functions (PDFs) is relatively well studied and explored, but its higher-dimensional counterpart, generalized parton distributions (GPDs), and tomographic imaging remain largely uncharted. Current and upcoming experiments aim to shed light on this structure \cite{khalek2022snowmass}. Complementing these experimental pursuits, first-principle investigations on the lattice play a crucial role.

Historically, lattice calculations primarily focused on determining different form factors, specifically moments of partonic distributions. However, addressing the full $x$-dependence of these distributions proved challenging due to the limitations imposed by the Euclidean spacetime metric. Around a decade ago, the groundbreaking concept of computing quasi-distributions marked a significant leap forward in overcoming these challenges \cite{ji2013}. This innovative approach involves determining these quasi-distributions using the lattice-calculable spatial correlations of a boosted hadron. Said distributions are later translated into relevant, physical (Minkowski) ones. This approach necessitates lattice observables that are not only renormalizable but also share the same infrared structure as their physical counterparts. Substantial theoretical and numerical achievements have been reached utilizing this approach, as detailed in various reviews, see e.g.\ \cite{cichy2019, ji2021,constantinou2021,cichy2022}.

There exists an alternative approach known as pseudo-distributions \cite{radyushkin2017, radyushkin2020}. This approach can be used on the same lattice data as quasi-distributions, but generates fundamentally different systematic effects. The crucial disparity arises from the manner in which they factorize into their physical distributions. Quasi-distributions perform this in momentum space, whereas pseudo-distributions do so in coordinate space. Despite this fundamental difference, both methodologies converge to the same physical distributions in the infinite momentum limit, provided all lattice-specific and other systematic effects are accounted for.

Previous work utilizing the pseudo-distribution approach concentrated on PDFs, see e.g.\ \cite{orginos2017lattice,joo2020,bhat2020,bhat2022}.
Extension of pseudo-PDFs to pseudo-GPDs was developed by Radyushkin \cite{radyushkin2019,radyushkin2020} and has not been subject to detailed lattice investigations so far.
In these proceedings, we share our results for the first exploration of the GPDs utilizing this approach.
As an example, we concentrate on the unpolarized $E$ GPD in the flavor non-singlet case ($u-d$).

\section{Theoretical background}
\subsection{Bare matrix elements}
The general form of the Euclidean bare matrix elements (MEs) for the vector case is
\begin{align}
F_V^\mu (z, P_f, P_i) = \langle N(P_f) | \bar\psi(z) \gamma^\mu W(0, z) \psi (0) | N (P_i) \rangle,
\end{align}
where $P_i$ and $P_f$ are the initial and final nucleon's four-momenta and $W(0,z)$ is the Wilson line taken in the $3$-direction and of length $z$. These matrix elements are obtained from a suitable ratio of three-point and two-point functions, see \cite{alexandrou2020} for details.
We work with one value of the source-sink separation, $t_s=10a$, and extract the MEs using the plateau method in the range of operator insertion times $3a \leq \tau \leq 7a$, which has been found to be robust at this level of statistical precision \cite{alexandrou2019lfo,alexandrou2020}.

\subsection{GPDs in coordinate space}
We calculate our bare MEs in asymmetric frames of reference, in which $P_f=(0,0,P_f^3)$ and $P_i=(-\Delta^1,-\Delta^2,P_i^3)$, where $\Delta^\mu$ is the momentum transfer and $P_i^3=P_f^3=P^3$ (zero-skewness case, $\Delta^3=0$).
The extraction of GPDs from such asymmetric frames was discussed extensively in \cite{bhattacharya2022}, where Lorentz-invariant amplitudes were introduced to parametrize MEs.
We recall that there are eight such linearly-independent amplitudes,
\begin{align}
    F_V^\mu (z, P_f, P_i) = \,\,&\bar{u}(P_f,\lambda')\Bigg[\frac{P^\mu}{m} A_1 + mz^\mu A_2 + \frac{\Delta^\mu}{m} A_3 + im\sigma^{\mu z} A_4  \\ &+ \frac{i\sigma^{\mu\Delta}}{m} A_5 + \frac{iP^{\mu} \sigma^{z\Delta} }{m} A_6 + imz^\mu\sigma^{z\Delta} A_7 + \frac{i \Delta^\mu \sigma^{z\Delta} }{m} A_8\Bigg]u(P_i,\lambda). \nonumber\label{amplitudes}
\end{align}
Additionally, we use four parity projectors $\Gamma_\kappa$ for each of the four gamma insertions $\gamma_\mu$ (see \cite{bhattacharya2022} for details), obtaining 16 different matrix elements. Eight of these are independent upon averaging equivalent contributions with reversed roles of $\Delta^1$ and $\Delta^2$. Thus, we obtain uniquely determined eight amplitudes.
Out of these, one can construct the Lorentz-invariant definition \cite{bhattacharya2022} of the $E$ GPD in coordinate space that we choose for this work,
\begin{align}
    E= -A_1 + 2A_5 + 2 z P^3 A_6.
\end{align}

\subsection{Renormalization and matching}
Once the coordinate-space Euclidean GPDs (pseudo-GPDs), $E(\nu,z)$ (where $\nu=P^3z$ is the Ioffe time), have been constructed from the amplitudes, our next step is to renormalize them. Bare MEs are riddled with Wilson-line-induced power divergences and standard logarithmic ones. Due to these being multiplicatively renormalizable \cite{ishikawa2017}, they can be eliminated using a ratio scheme, which is created using a combination of zero-momentum and local PDFs \cite{orginos2017lattice}. We employ the double ratio, defined as
\begin{align}
    \mathcal{E}(\nu, z) = \frac{E(\nu, z) / M(\nu, 0)}{M(0, z) / M(0, 0)}, \label{doubleratio}
\end{align}
where $M(\nu, z)$ is the unpolarized PDF. The double ratios are commonly referred to as  reduced Ioffe-time distributions (ITDs).
These reduced ITDs are Euclidean objects calculated at different scales $1/z$ and hence, they have to be evolved to a common scale and translated to Minkowski spacetime and the $\MSb$ renormalization scheme. We employ the one-loop matching \cite{radyushkin2018,zhang2018,izubuchi2018},
\begin{align}
    Q_E (\nu, \mu^2) = \mathcal{E}(\nu, \mu^2) - \frac{\alpha_s}{\pi} \int_0^1 du \, C(u, z, \mu^2) \left( \mathcal{E}(u\nu, \mu^2) - \mathcal{E}(\nu, \mu^2) \right), \label{matching}
\end{align}
where $C(u, z, \mu^2) = (C_F/2) \left( L(u) + B(u) \ln (z^2 \mu^2 e^{2\gamma_E + 1})/4 \right)$.
The function $B(u) = \frac{1 + u^2}{u - 1}$ handles the evolution of the ITD to a common scale $\mu$ (taken to be 2 GeV), and the function $L(u) = 4 \frac{\ln(1 - u)}{u-1} - 2(u - 1)$ matches the ITD to the light-cone and performs the scheme conversion.
The resulting quantity, $Q_E(\nu,\mu^2)$, is the light-cone $\MSb$-renormalized $E$ GPD in coordinate space.

\subsection{Reconstruction of the $x$-dependence}
The matched ITDs are related to the momentum-space GPDs, $q(x,\mu^2)$, by a Fourier transform, $q(x,\mu^2) = \int_{-\infty}^{\infty} d\nu\, e^{-i\nu x} Q_E(\nu, \mu^2)$.
However, the Fourier transform assumes an infinite range of continuous ITDs at all Ioffe times, while the lattice data are fundamentally discrete. This creates an inverse problem, which one can tackle in various ways \cite{karpie2019}.

In this work, we first consider the real and imaginary part of the ITDs separately,
\begin{align}
    \mathrm{Re} \, Q_E(\nu, \mu^2) &= \int_0^1 dx \cos(\nu x) q_v(x, \mu^2) \label{real_recon}\\
    \mathrm{Im} \, Q_E(\nu, \mu^2) &= \int_0^1 dx \sin(\nu x) q_{v2s}(x, \mu^2), \label{imag_recon}
\end{align}
where $q_v$ is the valence distribution, $q_v = q - \bar{q}$, and $q_{2vs} = q_v + 2\bar{q}$. We assume a fitting ansatz of the form $q(\mu, x) = N x^a (1 - x)^b$.
The valence distribution is normalized to $Q_E(0,\mu^2)$ and hence $N=Q_E(0,\mu^2)/B(a + 1, b+1)$ for this case, where $B$ is the Euler beta function.
For $q_{v2s}$, the normalization is not fixed  and $N$ is a fitting parameter, in addition to the parameters $a,b$ fitted in both cases. The fitting parameters are found by minimizing the $\chi^2$ function, $\chi^2 = \sum_{\nu = 0}^{\nu_{max}} ((Q_E(\nu, \mu^2) - Q_{E}^F (\nu, \mu^2))/\sigma_{Q_E(\nu, \mu^2)})^2$,
where the $Q_E^F(\nu, \mu^2)$ is the cosine/sine Fourier transform of the $x$-dependent fitting ansatz and $\sigma_{Q_E}$ is the error of $Q_E$.
The sum over Ioffe times is taken to be $\numax=P^3_{\rm max}\zmax$ and we discuss the choice of $\zmax$ below, while $P^3_{\rm max}$ is fixed by the maximum nucleon boost in our data.

\section{Lattice Setup}
We use one ensemble of $N_f = 2 + 1 + 1$ twisted mass fermions with a clover term, at a lattice spacing of $a = 0.093 \, \mathrm{fm}$ and a lattice volume of $T\cdot L^3/a^4=64 \cdot 32^3$. The pion mass is roughly twice the physical one, $m_\pi\approx260 \, \mathrm{MeV}$.

The pseudo-distribution approach is based on the same MEs as quasi-distributions.
Hence, we reuse the asymmetric-frame data of \cite{bhattacharya2022}, which are limited to a single value of $P^3=6\pi/L$.
To fully realize the potential of the pseudo-distribution approach, data at other values of $P^3$ can be used and additionally, zero-momentum MEs are needed.
Thus, we generated additional data corresponding to $P^3=\{0,1,2,4\}\cdot2\pi/L$.
The dominating computational effort concerned, obviously, the largest-momentum case, which allowed us to significantly enhance the covered range of Ioffe times.
The details of our setup are shown in Table \ref{tab:data}.

\begin{table}[h]
\centering
\begin{tabular}{|c|c|c|c|c|}
\hline
$P^3 [{\rm GeV}]$ & $N_{\mathrm{confs}}$ & $N_{\mathrm{kinematics}}$ & $N_{\mathrm{sourcepos.}}$ & $N_\mathrm{measurements}$ \\
\hline
\textcolor{red}{0.0} & \textcolor{red}{404} & \textcolor{red}{1} & \textcolor{red}{8} & \textcolor{red}{3,232} \\
\textcolor{red}{0.42} & \textcolor{red}{100} & \textcolor{red}{8} & \textcolor{red}{8} & \textcolor{red}{6,400} \\
\textcolor{red}{0.83} & \textcolor{red}{100} & \textcolor{red}{8} & \textcolor{red}{8} & \textcolor{red}{6,400} \\
1.25 & 269 & 8 & 8 & 17,216 \\
\textcolor{red}{1.67} & \textcolor{red}{404} & \textcolor{red}{8} & \textcolor{red}{32} & \textcolor{red}{103,424} \\
\hline
\end{tabular}
\caption{Details of our lattice setup. Given is the nucleon boost, the number of used gauge field configuratons, the number of kinematic setups, the number of source positions and the total number of measurements included in our statistics. New data generated for this work are highlighted in red and the ones in black correspond to data used previously for quasi-GPDs in \cite{bhattacharya2022}.}
\label{tab:data}
\end{table}

In these proceedings, we present results only for one value of the momentum transfer, $-t = 0.64 \, \mathrm{GeV}^2$, which is obtained from eight equivalent kinematic setups, i.e.\ momentum transfer 3-vectors $\vec{\Delta}$ taken as $(2,0,0)$, $(-2,0,0)$, $(0,2,0)$ and $(0,-2,0)$ at both signs of $P^3$.
As hinted above, we present only the case of the isovector $E$ GPD as an example.

\section{Results}

\begin{figure}[p]
    \centering
    \includegraphics[width=0.9\textwidth]{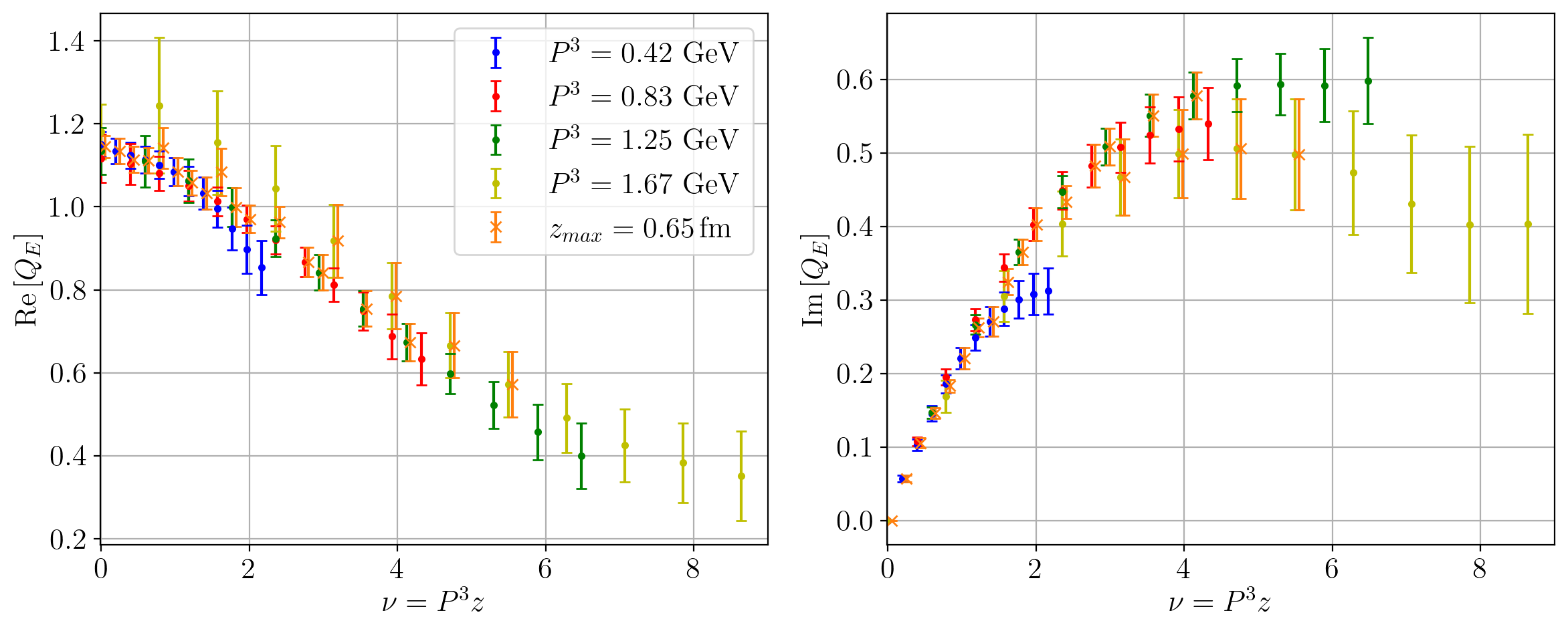}
    \caption{Matched $E$ ITDs at four nucleon momentum values (circles) and the averaged distribution (crosses), slightly shifted for better visibility.}
    \label{fig:matched}
    \centering
    \includegraphics[width=0.9\textwidth]{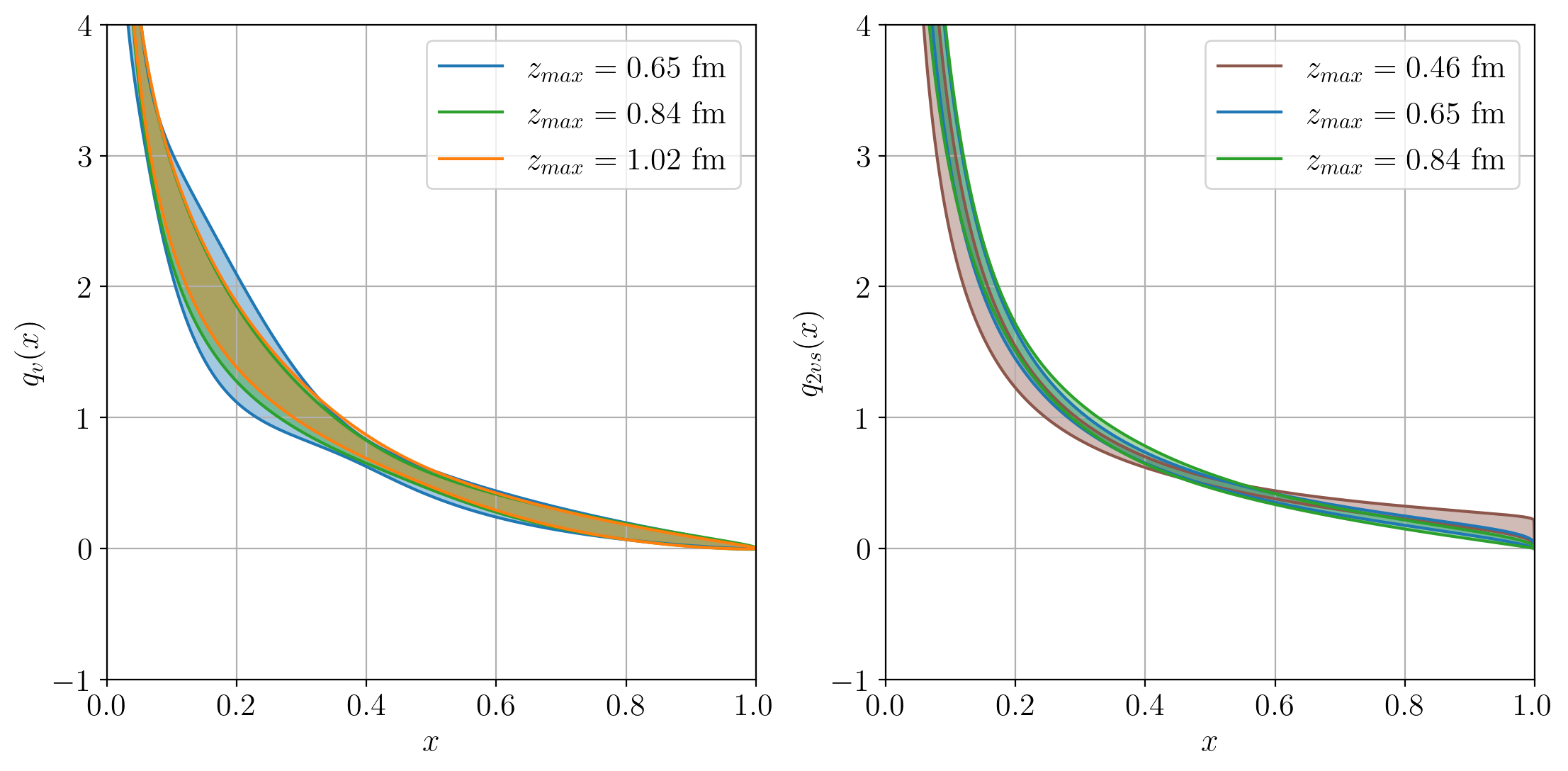}
    \caption{Left: $q_v(x)$, right: $q_{2vs} (x)$, reconstructed with different $\zmax$ values.}
    \label{fig:real_recon}
    \centering
    \includegraphics[width=0.9\textwidth]{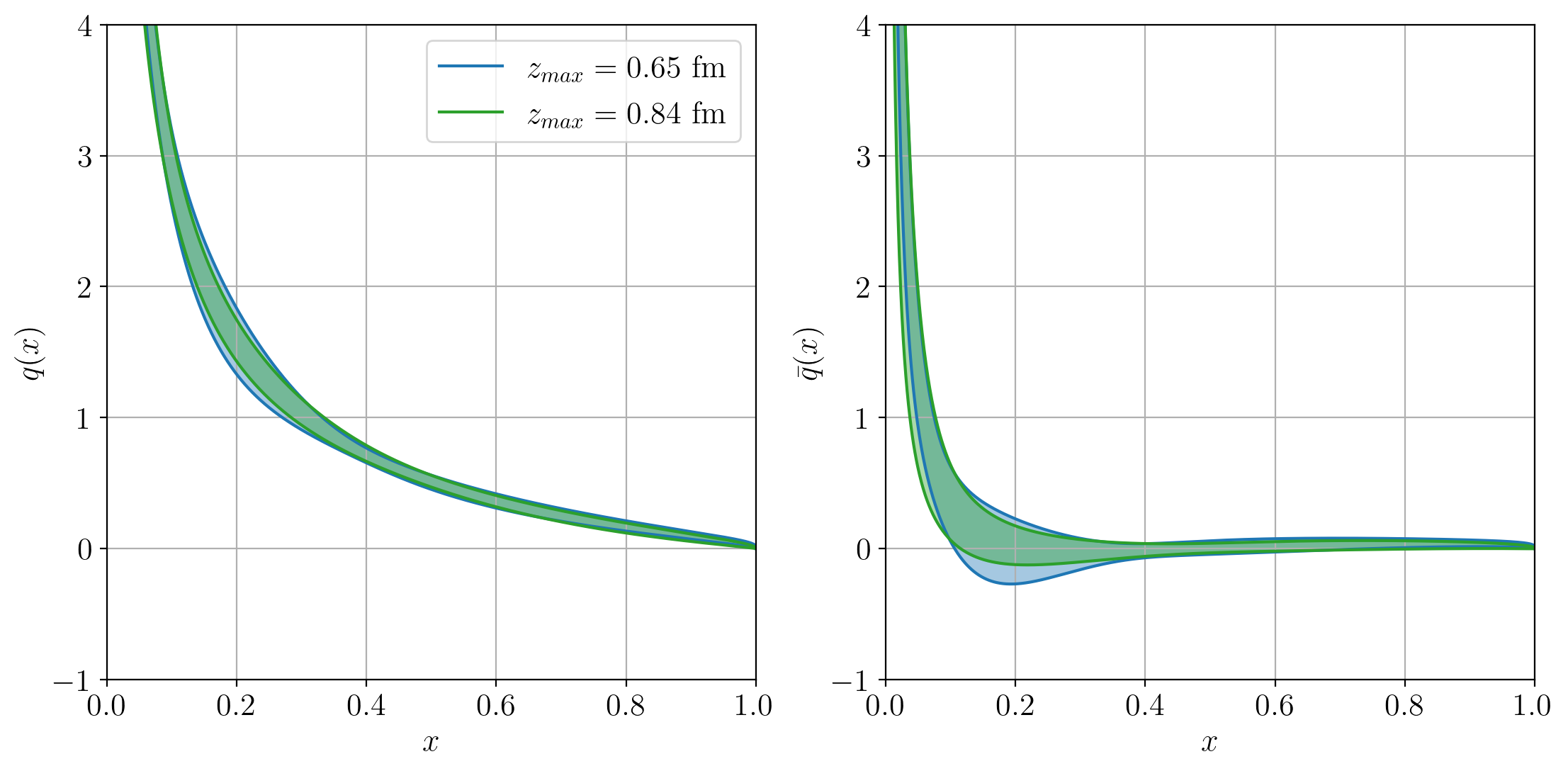}
    \caption{Left: $q(x)$, right: $\bar{q}(x)$, reconstructed with different $\zmax$ values.}
    \label{fig:imag_recon}
\end{figure}

In Figure \ref{fig:matched}, we present the matched ITDs calculated from (\ref{matching}). A given Ioffe time $\nu$ can be obtained from different combinations of $(P^3,z)$ and, in principle, ITDs from such combinations should be equivalent after they have been evolved to a common scale and matched. This holds, provided that the evolution/matching procedure is performed at a sufficiently small value of $z$.
Thus, the challenge is to find a suitable maximum value for $z$. We adopt the pragmatic criterion that $\zmax$ is the maximum value for which different $(P^3,z)$ pairs still lead to consistent values. Then, we average ITDs corresponding to the same Ioffe times and an example of this process is depicted in Figure \ref{fig:matched}, with circles pertaining to actual lattice data at a given boost and crosses to the averages over $(P^3,z)$ combinations (with data up to $\zmax=0.65$ fm entering the averages), where applicable.
The agreement between $(P^3,z)$ pairs persists until $z/a\approx10/8$ for the real/imaginary part.
Overall, we explore cases with $\zmax \in \{0.65, 0.84, 1.02\} \, \mathrm{fm}$ for the real part and for the imaginary part, displaying a tendency for larger effects, we look at more conservative set of values $\zmax \in \{0.47, 0.65, 0.84\} \, \mathrm{fm}$.

The final $x$-reconstructed distributions are presented in Figures \ref{fig:real_recon} (left: $q_v$, right: $q_{v2s}=q_v+2\bar{q}$) and \ref{fig:imag_recon} (left: $q=q_v+\bar{q}$, right: $\bar{q}$).
For the latter, we show results from common $\zmax$ values for the real/imaginary part data, although in principle, we could combine results from different $\zmax$ (we leave the analysis of the mixed cases for future work).
In any case, the obtained $\zmax$-dependence is relatively small, with all $\zmax$ values giving consistent $x$-dependent GPDs.
The obvious effect of decreasing $\zmax$ is the inflation of the error from missing parts of the Ioffe time range that constrain the behavior in $x$-space.

The results for the ``full'' distribution, $q=q_v+\bar{q}$, are qualitatively and to a large extent also quantitatively similar to the ones from the quasi-distribution approach \cite{bhattacharya2022}.
Larger differences are observed in the sea distribution, with quasi-GPDs leading to its significantly non-zero values already at $x\approx0.5$.
This points to the likely possibility that systematic effects evinced by both approaches are similar in the valence sector and enhanced for sea quarks.
However, the final conclusions can be reached only after a more systematic analysis, which is our current work in progress.
In our upcoming publication, we will report results for a wide range of momentum transfers for both $E$ and $H$ GPDs, with assessment of various systematic effects, such as the choice of the fitting ansatz.
Meanwhile, the presented example is the first exploration of pseudo-GPDs, with unambiguously encouraging results.
\vspace*{3mm}\\
\noindent \textbf{Acknowledgments}
NN, WC and KC acknowledge support by the National Science Centre grant OPUS no.\ 2021/43/B/ST2/00497.
SB was supported by the U.S. Department of Energy under Contract No. DE-SC0012704, and Laboratory Directed Research and Development (LDRD) funds from Brookhaven Science Associates.
MC acknowledges financial support by the U.S. Department of Energy, Office of Nuclear Physics, Early Career Award under Grant No.\ DE-SC0020405.
The work of AM was supported by the National Science Foundation under grant number PHY-2110472, and also by the U.S. Department of Energy, Office of Science, Office of Nuclear Physics, within the framework of the TMD Topical Collaboration.
FS was funded by by the NSFC and the Deutsche Forschungsgemeinschaft (DFG, German Research Foundation) through the funds provided to the Sino-German Collaborative Research Center TRR110 “Symmetries and the Emergence of Structure in QCD” (NSFC Grant No. 12070131001, DFG Project-ID 196253076 - TRR 110).
This research was supported in part by PLGrid Infrastructure (Prometheus supercomputer at AGH Cyfronet in Cracow).
Computations were also partially performed at the Poznan Supercomputing and Networking Center (Eagle supercomputer), the Interdisciplinary Centre for Mathematical and Computational Modelling of the Warsaw University (Okeanos supercomputer), at the Academic Computer Centre in Gda\'nsk (Tryton supercomputer), at facilities of the USQCD Collaboration, funded by the Office of Science of the U.S. Department of Energy and on resources of the Oak Ridge Leadership Computing Facility, which is a
DOE Office of Science User Facility supported under Contract DE-AC05-00OR22725.
The gauge configurations have been generated by the Extended Twisted Mass Collaboration on the KNL (A2) Partition of Marconi at CINECA, through the Prace project Pra13\_3304 ``SIMPHYS".
Inversions were performed using the DD-$\alpha$AMG solver~\cite{frommer2013} with twisted mass support~\cite{alexandrou2016}.

\end{document}